\documentclass[square]{ws-procs975x65}
\begin{document}
\title{General relativistic accretion with backreaction}
\author{Janusz Karkowski, Bogusz Kinasiewicz, Patryk Mach, Edward Malec}
\address{M. Smoluchowski Institute of Physics, Jagiellonian University, Reymonta 4, 30-059 Krak\'ow, Poland}
\author{Zdobys\l aw \'Swierczy\'nski}
\address{Pedagogical University, Podchor\c{a}\.zych 1, Krak\'ow, Poland}
\begin{abstract}
The spherically symmetric steady accretion of polytropic perfect
fluids onto a black hole is the simplest flow model that can demonstrate
the effects of backreaction. Backreaction keeps intact most of the
characteristics of the sonic point. For any such system the mass accretion
rate achieves maximal value when the mass of the fluid is 1/3 of the total
mass. Fixing the total mass of the system, one observes the existence of
two weakly accreting regimes, one overabundant and the other poor in fluid
content.
\end{abstract}
\bibliographystyle{ws-procs975x65}
\bodymatter
\section{Introduction}
Calculations of selfgravitating fluids onto a compact object are, in general, very difficult. So it is not suprising that this problem has been solved in only a few idealized cases. 
The spherical steady accretion of perfect fluids onto a Newtonian gravitational center was investigated by Bondi in 1952 \cite{bondi} and onto a Schwarzschild black hole by Michell \cite{michel} and others \cite{shapiro_teukolsky,kinasiewicz_lanczewski,kinasiewicz_mach}. 
The first fully general relativistic model taking into account the backreaction was dealt with by Malec \cite{malec}.
The effects of backreaction of selfgravitating fluids on a spherical black hole was examined in  \cite{karkowski}.
The influence of backreaction of steadily accreting gases on the stability has been studied in \cite{kinasiewicz}.   
In this paper we will briefly present the main results of \cite{karkowski}. 
\section{Formulation of the problem of quasistationary accretion}
Let us consider the spherically symmetric cloud of an ideal gas falling onto a non-rotating black hole. 
The general spherically symmetric line element is given by
\[ds^{2}=-N^{2}dt^{2}+\alpha dr^{2}+R^{2}d\theta^{2}+R^{2}sin^{2}{\theta}d\phi^{2},\]
where $N$, $\alpha$ and $R$ depend on the asymptotic time variable $t$ and the radius $r$. 
We assume the energy-momentum tensor of perfect fluid
$T^{\mu\nu} = (p + \varrho) u^\mu u^\nu + p g^{\mu \nu}$,
where $u^\mu$ denotes the four velocity of the fluid, $p$ is the pressure and $\varrho$ the energy density in the comoving frame.\\
The conservation of the energy-momentum tensor $\nabla_\mu T^{\mu \nu} = 0$ leads to the continuity equation  
$\partial_t \varrho = - N \mathrm{tr} K (\varrho + p)$
and to the relativistic version of the Euler equation 
$N \partial_R p + (\varrho + p) \partial_R N = 0.$
Here the extrinsic curvature $K_{r}^{r}=\partial_t \alpha / (2N\alpha)$ and $\mathrm{tr} K= N^{-1} \partial_t \ln \left( \sqrt{\alpha} R^2 \right)$ (for details see \cite{malec}). The quasilocal mass $m(R)$ is defined by $\partial_R m(R)=4\pi R^2\varrho$.
We will assume that the accretion is steady and the fluid satisfies the polytropic equation of state $p=K\varrho^{\Gamma}$ with constant $\Gamma\in(1,5/3]$.  
More precisely: 
\begin{itemize}
\item[{i.}] the accretion rate, defined as $\dot m = ( \partial_t - (\partial_t R)\partial_R)m(R)$ for the given areal radius $R$, is assumed to be constant in time; 
\item[{ii.}] the fluid velocity $U=(\partial_t R)/N$, energy density $\varrho$, sound velocity $a$ etc. should remain constant on the surface of fixed $R$: $(\partial_t - (\partial_t R) \partial_R) X = 0$, where $X = U, \varrho, a, \dots$
\end{itemize}
Strictly saying, a stationary accretion must lead to the increase of the central mass and of some geometric quantities. This in turn means that the notion "steady accretion" is approximate -- it demands the mass accretion rate is small and the time scale is short, so that the quasilocal mass $m(R)$ does not change significantly.\\
One can show \cite{malec} that the accretion rate is independent of the surface (characterised by a given $R$) for which it is calculated, i.e., $\partial_R \dot m=0$.
Let us now define a sonic point as such, where the length of the spatial velocity vector equals the speed of sound $|\vec U| = a$. In the Newtonian limit the above definition coincides with the standard requirement of the equality between the velocity of the fluid and the local sound speed. In the following we will denote by the asterisk all values referring to the sonic point.
\section{The importance of backreaction}
One of the two main results of \cite{karkowski} is the observation that significant information about the full system with backreaction can be obtained through the investigation of steady flows with the test fluid approximation.  
It appears that the characteristics $a_\ast^2$, $U_\ast^2$, $m_\ast/R_\ast$ of the sonic point {\it practically} do not depend, for a given $\Gamma$ and $a^2_\infty$, on the asymptotic energy density $\varrho_\infty$.
One can get all parameters describing the sonic point, with the exception of its location $R_\ast$ and mass $m_\ast$ simply from a related polytropic model with the test fluid having the same index $\Gamma$ and the same asymptotic speed of sound $a_\infty$.\\
\begin{figure} 
\begin{center}
\psfig{file=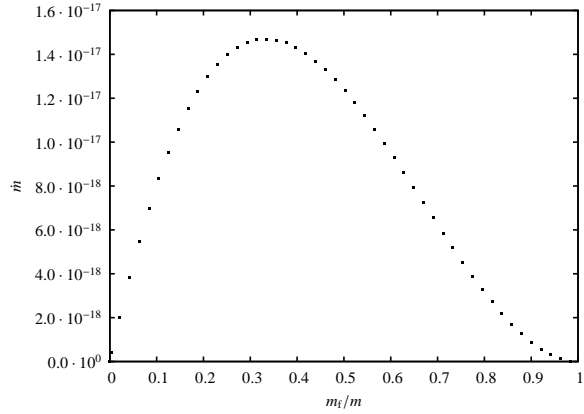,width=3in} 
\caption{The dependence of $\dot m$ on the ratio $m_f/m$ for $\Gamma=4/3$ and $a_{\infty}^2=0.1$.} 
\label{mdot} 
\end{center}
\end{figure}
The second main result come from investigation of the mass accretion rate \cite{karkowski} 
\[\dot{m}=m^3 x\frac{(1-x)^2}{\gamma}\pi\left(\frac{a_{\ast}^{2}}{a_{\infty}^{2}}\right)^{\frac{5-3\Gamma}{2(\Gamma -1)}}\left(1+\frac{a_{\ast}^2}{\Gamma}\right)\frac{1+3a_{\ast}^2}{a_{\infty}^3},\]
where $x=m_f /m$.
This expression clearly demonstrates that the mass accretion rate achieves a maximum at $m_f=m/3$ and tends to zero when $m_f\rightarrow 0$ and $m_f\rightarrow m$ (Fig. \ref{mdot}). The factor $1/3$ is universal -- independent of the parameters $R_\infty$, $\Gamma$ and $a_\infty$. In the test fluid approximation the situation is quite different -- the quantity $\dot{m}$ grows with $\varrho_\infty$.\\
The above result show the importance of the backreaction. We are convinced that this qualitative features demonstrated by the spherically symmetric model will also appear in the descriptions of accreting fluid onto a rotating black hole.
\section{Summary}
In conclusion, in the simple model of accretion with backreaction considered here, one can get all parameters describing the sonic point, with the exception of its location $R_\ast$ and mass $m_\ast$ simply from a related polytropic model with the test fluid having the same index $\Gamma$ and the same asymptotic speed of sound $a_\infty$.
The main result is that the mass accretion rate $\dot m$ achieves a maximum at $m_f/m_{BH}\approx 1/2$. Therefore, there exist two different regimes, $m_f/m_{BH}\ll 1$ and $m_f/m_{BH}\gg 1$, with low accretion. \\This paper has been partially supported by the MNII grant 1PO3B 01229.

\end{document}